\documentclass[prl,10pt,aps,floatfix,notitlepage,twocolumn,superscriptaddress]{revtex4-2}

\usepackage{aas_macros}
\usepackage{amsmath,amsfonts,amssymb}
\usepackage[colorlinks=true,citecolor=blue,urlcolor=blue]{hyperref}
\usepackage{graphicx}
\usepackage{mathrsfs}
\usepackage{xcolor}
\usepackage{mathtools, bm}
\usepackage{amsmath}
\usepackage{braket}
\usepackage{physics}
\usepackage{siunitx}
\usepackage{booktabs}
\usepackage{microtype} 
\usepackage{comment}
\usepackage{soul}

\makeatletter
\renewcommand\onecolumngrid{%
  \do@columngrid{one}{\@ne}%
  \def\set@footnotewidth{\onecolumngrid}%
  \def\footnoterule{\kern-6pt\hrule width 1.5in\kern6pt}%
}
\renewcommand\twocolumngrid{%
  \def\footnoterule{%
    \dimen@\skip\footins\divide\dimen@\thr@@
    \kern-\dimen@\hrule width.5in\kern\dimen@}%
  \do@columngrid{mlt}{\tw@}%
}
\makeatother

\newcommand{\mathd}{\mathrm{d}}

\usepackage{accents}
\newlength{\dhatheight}

\makeatletter
\def\bstctlcite{\@ifnextchar[{\@bstctlcite}{\@bstctlcite[@auxout]}}
\def\@bstctlcite[#1]#2{\@bsphack
	\@for\@citeb:=#2\do{%
		\edef\@citeb{\expandafter\@firstofone\@citeb}%
		\if@filesw\immediate\write\csname #1\endcsname{\string\citation{\@citeb}}\fi}%
	\@esphack}
\makeatother

\begin{document}
	
\title{Constraints on Dark Matter Structures around Gaia Black Holes}

\author{Nuno P.~Branco}
\email{nuno.branco@student.fis.uc.pt}
\affiliation{Univ Coimbra, Faculdade de Ci\^encias e Tecnologia da Universidade de Coimbra and CFisUC, Rua Larga, 3004-516 Coimbra, Portugal}

\author{Ricardo Z.~Ferreira}
\email{rzferreira@uc.pt}
\affiliation{Univ Coimbra, Faculdade de Ci\^encias e Tecnologia da Universidade de Coimbra and CFisUC, Rua Larga, 3004-516 Coimbra, Portugal}
\affiliation{Centro de Física das Universidades do Minho e do Porto (CF-UM-UP), Universidade do Minho, 4710-057 Braga, Portugal}

\author{Jo\~{a}o G.~Rosa}
\email{jgrosa@uc.pt}
\affiliation{Univ Coimbra, Faculdade de Ci\^encias e Tecnologia da Universidade de Coimbra and CFisUC, Rua Larga, 3004-516 Coimbra, Portugal}

\author{Rodrigo Vicente}
\email{r.l.lourencovicente@uva.nl}
\affiliation{Gravitation Astroparticle Physics Amsterdam (GRAPPA), University of Amsterdam, 1098 XH Amsterdam, The Netherlands}

\date{\today}

\begin{abstract}
\noindent
We demonstrate that Gaia's detection of stars on wide orbits around black holes opens a new observational window on dark matter structures---such as scalar clouds and dark matter spikes---predicted in a range of theoretical scenarios. Using precise radial velocity measurements of these systems, we derive state-of-the-art constraints on dark matter density profiles and particle masses in previously unexplored regions of parameter space. We also test the black hole hypothesis against the alternative of a boson star composed of light scalar fields.
\end{abstract}

\maketitle

\noindent{\textit{\textbf{Introduction.}---}}%
The Gaia mission is producing the most precise three-dimensional map of stellar positions and motions in the Milky Way~\cite{Gaia:2016}. Among its achievements is the discovery of three stellar-mass black holes (BHs) via their gravitational influence on companion stars~\cite{El_Badry_2022, El_Badry_2023, Panuzzo:2024}. 
Follow-up observations have determined the orbital parameters of these systems---the widest known binaries containing a stellar-mass BH~\cite{Panuzzo:2024}---with exceptional accuracy.

These systems offer a unique opportunity to probe theoretical dark matter (DM) structures expected to form around BHs. Examples include condensates of light bosons, arising through mechanisms such as superradiance~\cite{Brito:2015oca}, accretion~\cite{Hui:2019aqm, Clough:2019jpm, Bamber:2020bpu}, or dynamical capture~\cite{Budker:2023sex}, and dense DM spikes potentially linked to primordial BHs~\cite{Mack:2006gz, Eroshenko:2016yve, Boucenna:2017ghj, Adamek:2019gns, Carr:2020mqm}. Alternatively, the compact objects identified by Gaia might not be BHs at all, but boson stars (BSs) composed of light scalar fields~\cite{Kaup:1968zz, Ruffini:1969qy, Friedberg:1986tp, Lee:1991ax, Liebling:2012fv, Visinelli:2021uve, Seidel:1991zh}.

Such scenarios have been probed through independent observations.
Boson condensates are constrained by observations of highly spinning BHs~\cite{Arvanitaki:2010sy, Arvanitaki:2014wva, Brito:2014wla, Hoof:2024quk} and by the absence of gravitational-wave (GW)~\cite{Brito:2017wnc, Brito:2017zvb} or electromagnetic~\cite{Rosa:2017ury, Branco:2023frw, Ferreira:2024ktd} signals from the cloud dynamics or annihilation. Similarly, the existence of DM spikes has been probed via the non-detection of $\gamma$-ray signals from DM annihilation or decay~\cite{Boucenna:2017ghj, Adamek:2019gns, Carr:2020mqm}. Precise monitoring of the S2 star’s orbit by the GRAVITY Collaboration constrains the mass enclosed within its pericenter to a few thousand solar masses~\cite{GRAVITY:2024tth}, constraining both DM spikes~\cite{Lacroix:2018zmg} and light bosonic clouds~\cite{GRAVITY:2023cjt, GRAVITY:2023azi}.  Intriguingly, the anomalously rapid orbital decay of nearby low-mass X-ray binaries has been speculatively attributed to dynamical friction from DM spikes~\cite{Chan:2022gqd, Ireland:2024lye}. Looking ahead, GW observations will provide highly precise probes of BH environments through their effect on orbital phasing~\cite{Cardoso:2019rou, Cole:2022yzw, Cole:2022ucw, CanevaSantoro:2023aol, Bertone:2024rxe, Duque:2023seg, Vicente:2025gsg}.

In this Letter, we show that Gaia’s observations of wide BH-star binaries open a novel observational window into these scenarios. Using radial velocity data, we test two well-motivated BH environments---a scalar cloud and a DM spike---as well as the alternative hypothesis that the compact objects are BSs rather than BHs. We find that existing data is already remarkably sensitive to such DM structures.
We use natural units with $c=\hbar=1$.\\

\vspace{0.5mm}
\noindent{\textit{\textbf{Effect of DM on Stellar Orbits.}---}}%
The systems we consider consist of a BH, its surrounding DM structure, and a stellar companion. All Gaia BHs discovered so far are found in highly asymmetric binaries, with star-to-BH mass ratios in the range~$(0.02-0.1)$.
So, neglecting stellar tidal effects, they can be effectively modeled as a two-body problem: a star orbiting a composite BH+DM object.

The DM distribution sources an additional gravitational potential~$V_{\rm DM}$ satisfying the Poisson equation~$\nabla^2V_{\rm DM}=4\pi G \rho_{\rm DM}$, where~$\rho_{\rm DM}$ is the mass density of the DM bound to the BH. 
For spherically symmetric DM structures, the total orbital angular momentum is conserved, and the separation between the star and the BH+DM center of mass ($r\equiv|\bm{r}_\star-\bm{r}_\bullet|$), evolves according to
{\begin{equation} \label{eq: orbital EOM}
	\ddot{r}=\frac{M}{M_\bullet} \left[\left(\frac{M}{ M_\bullet}\right)\frac{L^2}{r^3}-\frac{G M_{\rm BH}}{r^2}-\frac{\mathd V_{\rm DM}}{\mathd r}\right]\,,
\end{equation}}
where~$M_\bullet\equiv M_{\rm BH}+M_{\rm DM}$ is the BH+DM mass,~$M\equiv M_\bullet+M_\star$ is the total system mass, and~$\bm{L}\equiv (M_\bullet/M)(\bm{r}\times \dot{\bm{r}})$ is the orbital angular momentum per unit stellar mass. The orbital energy and angular momentum are fixed through the energy conservation equation 
\begin{equation}
    \frac{\dot{r}^2}{2} + V_{\text{eff}}(r) =\mathrm{const.}\,,
    \label{eq: 4-velocity constraint}
\end{equation}
by prescribing $\dot{r}=0$ at $r_\text{max}=a(1+e)$ and $r_\text{min}=a(1-e)$, with semi-major axis $a$, and eccentricity $e$; $\mathd V_{\text{eff}}(r)/\mathd r$ coincides with the right-hand side of Eq.~\eqref{eq: orbital EOM}. In the perturbative regime, the stellar motion can be described as an apsidally precessing ellipse.\\

\vspace{0.5mm}
\noindent{\textit{\textbf{Gaia BHs and Data Analysis.}---}}%
Gaia observations combining astrometric, photometric, and spectrometric (radial velocity) data have revealed three stars orbiting stellar-mass BHs, designated Gaia BH1, BH2, and BH3~\cite{El_Badry_2022, El_Badry_2023, Panuzzo:2024}.
These binaries are wide enough that the stars evolve in the weak gravitational field of the BHs, and so relativistic effects are negligible. 
In the absence of additional forces, the orbits are Keplerian ellipses described by the standard orbital parameters---period \(P\), eccentricity \(e\), argument of periastron \(\omega\), inclination \(i\), stellar mass \(M_{*}\), semi-major axis \(a\), and periastron time \(T_p\)---extracted from fits to Gaia's astrometry and radial velocity data (see~\cite{El_Badry_2022, El_Badry_2023, Panuzzo:2024} for further details). 

\begin{table}[t]
    \centering
    \caption{Orbital parameters from Gaia's astrometry data for the three Gaia BH systems~\cite{El_Badry_2022, El_Badry_2023, Panuzzo:2024}.
    }
    \label{tab:bh_parameters}
    \begin{tabular}{lccc}
        \toprule
        Parameter & BH1 & BH2 & BH3 \\
        \midrule
        $M_{\star}$ ($M_\odot$) & $0.93 \pm 0.05 $ & $1.07 \pm 0.19 $ & $0.76 \pm 0.05 $ \\
        $P$ (days)            & $185.8 \pm 0.3$             & $1300 \pm 26$             & $4194.7 \pm 112.3$            \\
        $e$                   & $0.49 \pm 0.07$              & $0.515 \pm 0.01$            & $0.726 \pm 0.006$             \\
        $\omega$ (deg)        & $-10.3  \pm 11.6$              & $131.2 \pm 1.6$             & $77.8  \pm 0.7$             \\
        i (deg)     & $121.2 \pm 2.8$              & $35.7 \pm 0.7$              & $110.66 \pm 0.11$            \\
        $a$ (AU)               & $1.40 \pm 0.01$ & $5.05 \pm 0.12$ & $16.17 \pm 0.27$ \\
     $T_p$ (JD)               & $2457377 \pm 6$ & $2457438.4 \pm 3.1$ & $2458177.3 \pm 1.0$ \\
        \bottomrule
    \end{tabular}
\end{table}

If a DM overdensity surrounds the BH, its additional gravitational potential can induce measurable deviations from these Keplerian orbits. We therefore use currently available data to constrain such deviations for three DM configurations, described in the following sections. 
For each case, we simulate stellar orbits by integrating Eq.~\eqref{eq: orbital EOM} with initial conditions set by Eq.~\eqref{eq: 4-velocity constraint} at the periastron.
From these trajectories we compute the radial velocity as a function of time and compare directly with Gaia measurements supplemented by radial velocity follow-up data~\cite{El_Badry_2022, El_Badry_2023, Panuzzo:2024, Nagarajan_2024}.

All orbital parameters constrained by Gaia astrometry are fixed to their measured values~\cite{El_Badry_2022, El_Badry_2023, Panuzzo:2024} (see Tab.~\ref{tab:bh_parameters}), except for the BH masses (estimated by Gaia as $M_{\text{BH1}}\approx9.6 M_\odot$, $M_{\text{BH2}}\approx8.9 M_\odot$, and $M_{\text{BH3}}\approx33 M_\odot$), which are left free due to their strong degeneracy with the DM parameters.
Together with the center-of-mass radial velocity~$\gamma$, we treat them as part of a nuisance parameter set~$\bm{\nu}\equiv(M_{\text{BH}},\gamma)$.

\begin{figure}[t!]
    \centering
    \includegraphics[width=0.48\textwidth]{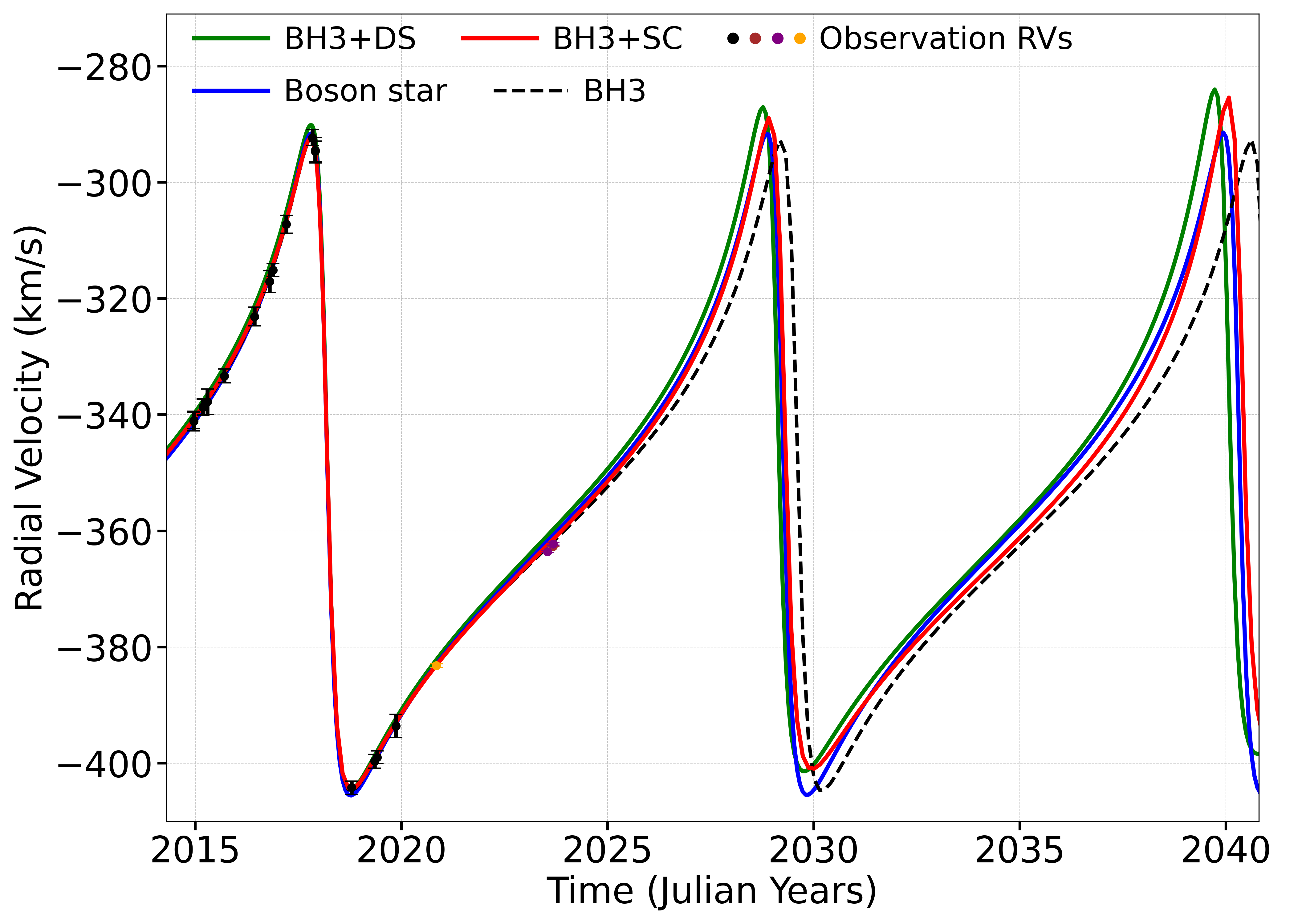}
    \caption{\small
    Observed stellar radial velocities in the Gaia BH3 system from Gaia epoch data (black symbols), HERMES (purple), SOPHIE (red), and UVES (yellow)~\cite{Panuzzo:2024, Raskin_2011, PERRUCHOT:2011, Dekker22}, compared with our models. The dashed black curve shows the best-fit Keplerian orbit. Alternative scenarios include: a scalar cloud ($\mu=10^{-15}\,\text{eV}$, $q=0.05$; red), a boson star ($\mu=4.2\times10^{-15}\,\text{eV}$; blue), and a DM spike ($\rho(a)=6\times10^{12}\,\mathrm{GeV\,cm^{-3}}$, $s=-9/4$; green). 
    These non-Keplerian cases would be excluded at the $2\sigma$ level by 2035, assuming data collection continues at the same rate.
    }
    \label{fig:RVplots}
\end{figure}

Constraints on the DM parameter set $\bm{\theta}_{\rm DM}$ are obtained from the profile likelihood ratio~\cite{pdg2024_statistics, Sprott:2000}, with $\chi^2(\bm{\theta}_{\rm DM},\bm{\nu})\equiv \sum_{i=1}^N [v_i^{\rm obs}-v_i^{\rm DM}(\bm{\theta}_{\rm DM},\bm{\nu})]^2/\sigma_i^2$, where \(v^{\mathrm{obs}}_{i}\) and \(\sigma_{i}\) are the observed radial velocity and its uncertainty at time \(t_i\), and \(v^{\mathrm{DM}}_{i}\) is the predicted velocity from Eq.~\eqref{eq: orbital EOM}.
The test statistic~$\Delta\chi^{2}(\bm{\theta}_{\rm DM})\equiv \min_{\bm{\nu}} \chi^2 - \min_{(\bm{\theta}_{\rm DM},\bm{\nu})}\chi^2$ follows a chi-squared distribution with $\dim \bm{\theta}_{\rm DM}$ degrees of freedom under Wilks' theorem~\cite{Wilks:1938dza, Rolke:2004mj}. This allows us to identify regions of parameter space excluded at the 68\% and 95\% confidence levels ($\simeq 1\sigma$ and $2\sigma$). Figure~\ref{fig:RVplots} compares the observed radial velocities of Gaia BH3 with the best-fit Keplerian orbit (no DM) and with representative predictions for scalar clouds, boson stars, and DM spikes. All examples shown exhibit the retrograde precession characteristic of extended DM profiles \cite{10.1119/1.14287, Chan_2022, refId0}.  \\ 

\noindent{\textit{\textbf{Scalar Clouds.}---}}%
One possible DM structure around BHs is a cloud of light scalar particles occupying the bound states of a \textit{gravitational atom}~\cite{Baumann:2019eav}. For highly spinning BHs, superradiance can populate axially symmetric states co-rotating with the BH~\cite{Brito:2014wla, Arvanitaki:2014wva, Brito:2015oca}. More recently, it was shown that attractive quartic self-interactions of axion-like particles (ALPs) can trigger stimulated capture around stars, populating the \emph{spherically symmetric} ground-state~\cite{Budker:2023sex}. 

BHs can similarly act as seeds for stimulated capture, allowing the condensation of DM particles into more compact configurations. The growth is most efficient in the regime~$\alpha\equiv G M_{\rm BH} \mu \gtrsim \sigma_v/(2\pi)$, where~$\sigma_v$ is the local velocity dispersion of unbound DM particles and $\mu$ is the mass of the scalar field. In this case, the capture timescale is~\cite{Budker:2023sex}
{\small
\begin{equation*}
    \frac{\tau_{\rm grow}}{10^9 \,{\rm yr}}\simeq \left[\frac{\mu}{10^{-15}{\rm eV}}\right]^3 \left[\frac{f_a}{10^{13}{\rm GeV}}\right]^4\left[\frac{10^{6}\,\mathrm{GeV/cm^3}}{\rho_\infty}\frac{\sigma_v}{1\,\mathrm{km/s}}\right]^2\,,
\end{equation*}}
with~$f_a$ the axion-like decay constant (which parametrizes the strength of the self-interactions) and $\rho_\infty$ the ambient DM density. Notably, this timescale is independent of the seed BH mass.
Growth competes with absorption by the BH, which occurs on a timescale $\tau_{\rm abs}\simeq 10^{11}\, \mathrm{yr} \left(10^{-15}\,{\rm eV}/\mu\right)\left(10^{-4}/\alpha\right)^5$~\cite{Detweiler:1980uk, Cardoso:2022nzc}.

\begin{figure}[t!]
    \centering
\includegraphics[width=0.48\textwidth]{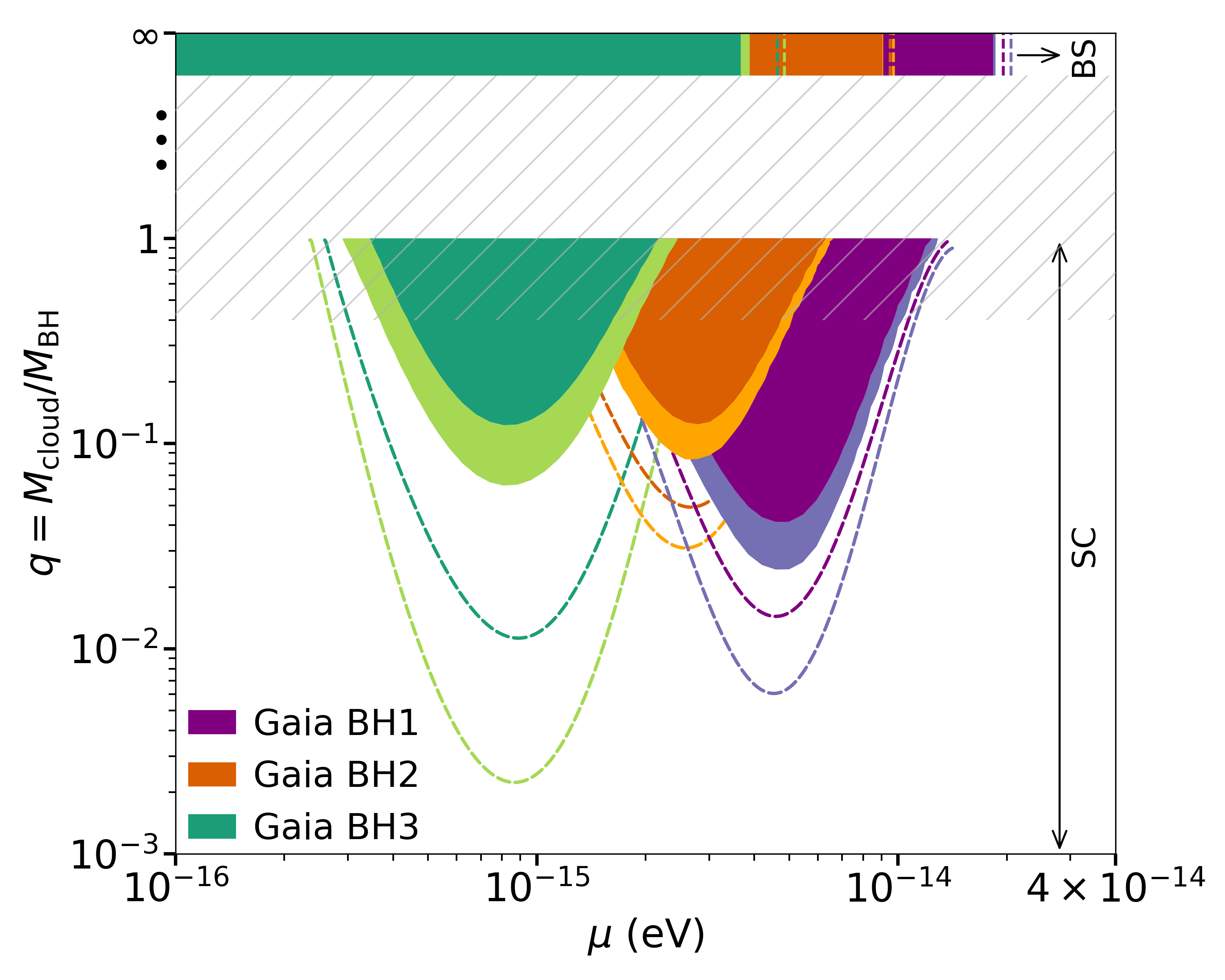}
    \caption{\small
    Excluded parameter space for scalar clouds (SC) and boson stars (BS) from radial velocity data. Filled bands show $2\sigma$ (dark) and $1\sigma$ (light) limits for Gaia BH1 (purple), BH2 (orange), and BH3 (green); dashed contours indicate projections assuming similar data acquisition rates until 2035. The hatched region denotes where scalar self-gravity becomes important and the SC model ceases to be valid. The limit $q\to\infty$ corresponds to a purely self-gravitating BS.
    }
    \label{fig: finalregion}
\end{figure}

For our analysis, we take a model-independent approach and consider a generic spherically symmetric ground-state scalar cloud (SC). We parameterize it by the cloud-to-BH mass ratio~$q\equiv M_\text{DM}/M_\text{BH}$, with gravitational potential 
\begin{gather}
	V_{\rm SC}=-\frac{G M_{\rm DM}}{r}\left[1-e^{- \frac{2\alpha^2 r}{G M_{\rm BH}}}\left(1+ \frac{ \alpha^2r}{G M_{\rm BH}}\right)\right].
    \label{eq: Vdm}
\end{gather}
Figure~\ref{fig: finalregion} shows the regions of parameter space excluded at the $1\sigma$  and $2\sigma$ levels by radial velocity observations, from the likelihood analysis described above. We restricted the analysis to~$q<1$ and hatched the region~$q\gtrsim0.4$, where self-gravity of the field becomes important and Eq.~\eqref{eq: Vdm} no longer applies. Also shown are projected sensitivities assuming continued radial velocity monitoring of the Gaia BH systems at the current rate until 2035.

Our results show that radial velocity observations of Gaia's wide BH binary systems primarily constrain very light scalar field masses, for which the SC radius (the gravitational analogue of the Bohr radius) is comparable to the orbital size. Heavier particles remain unconstrained, since the associated compact SCs lie completely inside the orbit and are therefore indistinguishable from a BH. Among the known systems, Gaia BH1 probes the smallest cloud to BH mass-ratios, reaching~$q \simeq 3\times 10^{-2}$. Remarkably, continued monitoring of Gaia BH3 until 2035 could probe down to $q \sim 10^{-3}$.

Observations of these systems probe particle masses in the range $[10^{-16},10^{-14}]\,\mathrm{eV}$, a region not constrained by superradiance bounds from BH spin measurements; for stellar-mass BHs, these correspond to~$\alpha \lesssim 10^{-3}$, for which superradiance is inefficient on cosmological timescales. For stimulated capture, we verified that the parameter space probed is physically relevant: the cloud can grow faster than it is absorbed by the BH, while self-interactions remain weak enough for perturbation theory to apply.
A BH with $\sim 30 M_\odot$ can efficiently seed a SC of particles with $\mu\simeq 10^{-15} \,\mathrm{eV}$, which can reach a mass-ratio as large as $q\gtrsim 10^{-3}$~\footnote{The perturbative results are valid for mass-ratios~\cite{Budker:2023sex} $q\leq 2\times10^{-3}\Bigl[\frac{M_{\rm BH}}{10M_{\odot}}\Bigr]\Bigl[\frac{10^{-15}\mathrm{eV}}{\mu}\Bigr]\Bigl[\frac{10^{-4}}{\alpha}\Bigr]\Bigl[\frac{f_a}{10^{13}\,\mathrm{GeV}}\Bigr]^{2}$.} in an environment with ambient DM density $\rho_\infty\sim 10^5\,\mathrm{M_\odot/pc^3}$ and velocity dispersion $\sigma_v\sim 10\,\mathrm{km/s}$---conditions that could plausibly occur at the core of star clusters \cite{Sharma:2024ndf}. Note that Gaia BH3 is potentially part of the ED-2 stream, which is likely the remnant of a tidally disrupted globular cluster \cite{Panuzzo:2024}.\\

\noindent{\textit{\textbf{Boson Stars.}---}}%
Another possibility is that the dark components in the Gaia binaries are not BHs but self-gravitating DM structures~\cite{Tkachev:1986tr, Hogan:1988mp}. A prime candidate is the scalar BS, a horizonless, self-gravitating scalar configuration that effectively corresponds to the limit of large cloud-to-BH mass-ratio, $q \to \infty$ (i.e.~$M_{\rm BH}\rightarrow 0$).

If the U(1) global symmetry is broken after inflation, small-scale ALP perturbations can collapse into miniclusters at matter–radiation equality, which may later form axion stars. Notably, in this scenario, the axion star mass function peaks around $\sim10 M_\odot$ for $\mu\sim10^{-16}\,\mathrm{eV}$, with up to $\mathcal{O}(0.4)$ of all axion DM bound in such objects~\cite{Chang:2024fol}.

The gravitational potential of a BS, $V_{\rm BS}$, is usually obtained numerically; for example, an 11th-order polynomial fit reproduces the solution at the percent level~\cite{Annulli:2020lyc}.
We find, however, that the much simpler form
\begin{equation}
 V_{\rm BS}=-\frac{G M_{\text{DM}}}{r}\tanh{\left[\frac{G M_{\text{DM}} \mu^2 r}{\pi}\right]}\, ,
\end{equation}
fits the numerical solution equally well.
 
Using this potential, we perform the profile likelihood analysis and show in Fig.~\ref{fig: finalregion} the $1\sigma/2\sigma$-excluded regions. As in the case of SCs, significant deviations appear once the BS radius becomes comparable to the orbital size. This constrains the BS parameter space allowed by radial velocity data, requiring $\mu \gtrsim 2\times10^{-14}\,\mathrm{eV}$ for Gaia BH1, $\mu \gtrsim 10^{-14}\,\mathrm{eV}$ for BH2, and $\mu \gtrsim 4\times10^{-15}\,\mathrm{eV}$ for BH3. These limits exclude the most physically motivated region for BS formation from ALP miniclusters, expected if the U(1) global symmetry is broken after inflation \cite{Chang:2024fol}.\\

\noindent{\textit{\textbf{Dark Matter Spikes.}---}}%
Another well-motivated BH environment is a DM spike~\cite{Mack:2006gz,Ireland:2024lye, Eroshenko:2016yve, Boucenna:2017ghj, Adamek:2019gns, Carr:2020mqm}. While spikes around astrophysical stellar-mass BHs are expected to be relatively dilute, primordial BHs can host much denser spikes (e.g., via accretion). We model the density as a single power law, $\rho_{\text{spike}}(r) = A \, r^{s}$ with $-3<s \leq 0$; this brackets common cases in the literature~\cite{Boudaud:2021irr} (broken power laws interpolating between $s=0$ and $s=-9/4$ can also arise if the BH forms before kinetic decoupling~\cite{Eroshenko:2016yve, Adamek:2019gns, Ireland:2024lye}).
The enclosed mass and gravitational potential are respectively given by $M_{\text{DM}}(r) = 4\pi A\,  r^{3+s}/(3+s)$ and
\begin{equation}
    V_{\text{spike}} =\frac{G M_{\text{DM}}(r)}{(2+s)r}\,, \qquad  s\neq -2 \, ;
\end{equation}
the case $s=-2$ yields a logarithmic potential which we do not treat further.

\begin{figure}[t!]
    \centering
    \includegraphics[width=0.48\textwidth]{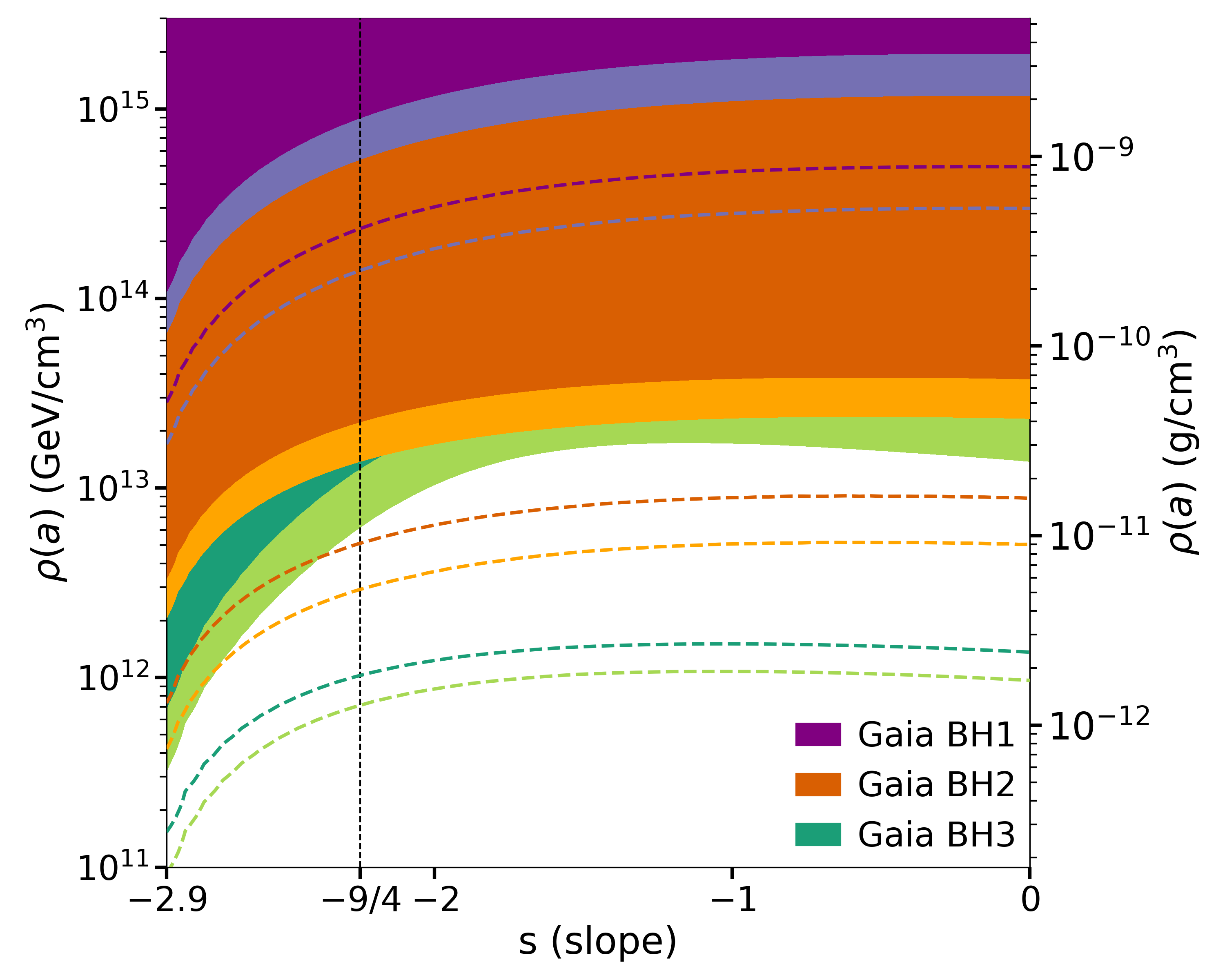}
    \caption{\small
    Excluded parameter space for DM density spikes around the Gaia black holes. Filled bands show $2\sigma$ (dark) and $1\sigma$ (light) limits on the spike density at the stellar semi-major axis, $\rho(a)$, for Gaia BH1 (purple), BH2 (orange), and BH3 (green). Dashed contours are projections assuming similar data acquisition rates until 2035. Regions above the colored bands are excluded at the indicated confidence level.
    }
    \label{fig: finalregionuniform}
\end{figure}

We perform the profile likelihood analysis in terms of the spike slope $s$ and the density at the semi-major axis, $\rho(a)$. Figure~\ref{fig: finalregionuniform} shows the $1\sigma/2\sigma$-excluded regions. The tightest constraints arise from Gaia BH3, excluding $\rho(a)\gtrsim 4\times10^{11}\,\mathrm{GeV\,cm^{-3}}$, in tension with primordial BH spike models where DM kinetically decouples before $T\sim 10\,\mathrm{keV}$ (cf. Fig.~1 of~\cite{Ireland:2024lye}). These limits are several orders of magnitude stronger than bounds from GW environmental effects ($\lesssim 10^{25}\,\mathrm{GeV\,cm^{-3}}$~\cite{CanevaSantoro:2023aol}) and depend only weakly on the slope $s$.\\

\noindent{\textit{\textbf{Discussion.}---}}%
Our results highlight Gaia’s ability not only to discover BHs and measure their orbital parameters through astrometry, but---once supplemented by radial velocity follow-up---to probe their local environments and thereby test the nature of DM. The constraints shown here rely primarily on radial velocity data, obtained both by Gaia’s spectrograph and by dedicated ground-based facilities (e.g., HERMES, SOPHIE, UVES, ESPRESSO).

With the forthcoming Gaia DR4 release, we expect a significant increase in the number of identified BH–star binaries, some with different orbital parameters that will broaden the scope of DM tests. 
Systems with high eccentricity are particularly valuable, since they allow the companion star to probe a wide range of radii within a single orbit. Shorter orbital periods also enhance sensitivity to departures from Keplerian motion, namely the induced retrograde precession~\cite{10.1119/1.14287, Chan_2022, refId0}, providing a promising discriminant for local DM densities. In this regard, continued high-cadence radial velocity monitoring---such as the ESPRESSO follow-up of Gaia BH1~\cite{Nagarajan_2024}---will be crucial.

The BHs identified by Gaia may also represent the same population probed by LIGO–Virgo–KAGRA. These observations are complementary: GW signals reveal the strong-field, dynamical regime of BH mergers where any surrounding DM structure is expected to be at least partially disrupted~\cite{Kavanagh:2020cfn}, while Gaia combined with radial velocity follow-up targets long-lived wide binaries, where DM configurations can persist. 

This synergy makes Gaia and future Gaia-like surveys unique laboratories for both BH astrophysics and DM physics, and our results motivate dedicated radial velocity campaigns and further theoretical work to fully exploit this potential.

\section*{Acknowledgments}

NPB is supported by the FCT fellowship PRT/BD/154617/2022. RV gratefully acknowledges the support of the Dutch Research Council
(NWO) through an Open Competition Domain Science-M grant, project number OCENW.M.21.375.
This work was partially supported through national funds by FCT (Fundação para a Ciência e Tecnologia, I.P.), with DOI identifiers 10.54499/2023.11681.PEX, 10.54499/UIDB/04564/2020, 10.54499/UIDP/04564/2020 
and by the projects with DOI identifiers 10.54499/2024.00249.CERN and 10.54499/2024.00252.CERN funded by measure RE-C06-i06.m02 – ``Reinforcement of funding for International Partnerships in Science, Technology and Innovation'' of the Recovery and Resilience Plan - RRP, within the framework of the financing contract signed between the Recover Portugal Mission Structure (EMRP) and the Foundation for Science and Technology I.P. (FCT), as an intermediate beneficiary.

\noindent 

\bibliography{mybibfile}
\end{document}